\documentclass[conference]{IEEEtran}
\IEEEoverridecommandlockouts
\usepackage{cite}
\usepackage{subcaption}
\usepackage{amsmath,amssymb,amsfonts}
\usepackage{algorithmic}
\usepackage{graphicx}
\usepackage{textcomp}
\usepackage{xcolor}
\def\BibTeX{{\rm B\kern-.05em{\sc i\kern-.025em b}\kern-.08em
    T\kern-.1667em\lower.7ex\hbox{E}\kern-.125emX}}
\begin{document}
\title{On Quantum-Enhanced LDPC Decoding for Rayleigh Fading Channels\\
}

\author{
\IEEEauthorblockN{Utso Majumder}
\IEEEauthorblockA{\textit{Jadavpur University}\\
Kolkata, India\\
utsomajumder@gmail.com}
\and
\IEEEauthorblockN{Aditya Das Sarma}
\IEEEauthorblockA{\textit{Jadavpur University}\\
Kolkata, India\\
aditya.41200@hotmail.com}
\and
\IEEEauthorblockN{Vishnu Vaidya$^{\dagger}$ \thanks{$^\dagger$ The author was with TCS Research and Innovation when the work was carried out}}
\IEEEauthorblockA{\textit{University of California}\\
San Diego, USA\\
vishnuvaidya98@gmail.com}
\and
\IEEEauthorblockN{M Girish Chandra}
\IEEEauthorblockA{\textit{TCS Research}\\
Bangalore, India\\
m.gchandra@tcs.com}
}
\maketitle
\begin{abstract}
Quantum and Classical computers continue to work together in tight cooperation to solve difficult problems. The combination is thus suggested in recent times for decoding the Low Density Parity Check (LDPC) codes, for the next generation Wireless Communication systems. In this paper we have worked out the Quadratic Unconstrained Binary Optimization (QUBO) formulation for Rayleigh Fading channels for two different scenarios- channel state fully known and not known. The resultant QUBO are solved using D-Wave 2000Q Quantum Annealer and the outputs from the Annealer are classically postprocessed, invoking the notion of diversity. Simple minimum distance decoding of the available copies of the outputs led to improved performance, compared to picking the minimum-energy solution in terms of Bit Error Rate (BER). Apart from providing these results and the comparisons to fully classical Simulated Annealing (SA) and the traditional Belief Propagation (BP) based strategies, some remarks about diversity due to quantum processing are also spelt out. 
\end{abstract}
\vspace{5pt}
\section{Introduction}

A key issue in communication over a wireless channel is the protection of the transmitted message against corruption by the noise in channel. This done by adding some redundancy along with the transmitted message that ensures that even in the presence of noise or interference, sufficient information reaches the receiver to allow the recovery of the transmitted message. This process of proofing the message by adding redundancy in some form, is known as channel coding. Some of the most popular channel codes in practical use today are the \textit{Turbo Codes}, and \textit{Low-Density Parity Check Codes} (or \textit{LDPC Codes}). LDPC codes, first introduced by Gallagher in 1962, can perform close to Shannon limit. But, the decoding of LDPC coded messages consume significant computation. Currently, the most popular classical method to decode LDPC coded messages is the Belief Propagation (BP) algorithm and its variants. In Kasi et al 
[6], the authors introduce a quantum annealer-based decoder solution, called the Quantum Belief Propagation (QBP), formulating the decoding problem as a Quantum Unconstrained Binary Optimization (QUBO) problem and use commercial quantum annealer (D-Wave) to perform the decoding. They considered Bipolar Phase Shift Keying (BPSK) symbols for the transmitted message and restricted to Additive White Gaussian Noise (AWGN) noise for the channel. In our earlier work 
[10] [11], we augmented the QBP with a classical post-processing step, enhancing the decoder performance. 

In this paper, we consider the more realistic setting of Rayleigh fading channels and suitable modifications towards extending the existing AWGN formulation are presented. Both coherent (Channel State Information 
is known) and non-coherent (unknown CSI) scenarios are considered. To the best of our knowledge, these extensions are novel. Further, carrying over from our previous proposition, we have explored the advantages of the post-processing in each of the said cases. The notion of receive diversity we spelt out in is also examined to certain detail in this paper considering both Rayleigh fading and AWGN channels.

For brevity and to avoid repetitiveness, in this paper, we have avoided the capturing of the well-known remarks on LDPC, Sum-Product or BP Decoding algorithm, Quantum Computing and Noisy Intermediate Scale Quantum (NISQ) hardware, and others. 
See 
[5] [1] [4] and the references there in for LDPC and the associated decoding. For Quantum Computing and NISQ, 
[9], [8] and the references there in can be consulted.

This paper is organized as follows: In Section II, just enough supporting preliminaries is covered. Section III address the QUBO formulations of LDPC decoding for Rayleigh fading channels; Section IV provide results, where, classical post processing is also adopted and the related discussions; Conclusions occupy Section V.  

\section{Some Essential Background}
\subsection{Low-Density Parity Check (LDPC) Coding}\label{AA}
A ($N, K)$ Low-Density Parity Check (LDPC) Code is a linear block code constructed with a sparse parity matrix $\mathbf{H}$ (therefore the \textit{Low-Density} in name LDPC). 
\begin{equation}
\label{eqn:3}
    \mathbf{H}\ =\ [h_{ij}]_{N \times K}
\end{equation}
It is called a $(d_b, d_c)$-regular code if every bit node $b_i, i \in [K]$ participates in $d_c$ checks and every check $c_i, i \in [N]$ is applied to $d_b$ bit nodes forming a \textit{check constraint}. 
\begin{figure}
    \centering
    \includegraphics[width=.99\columnwidth]{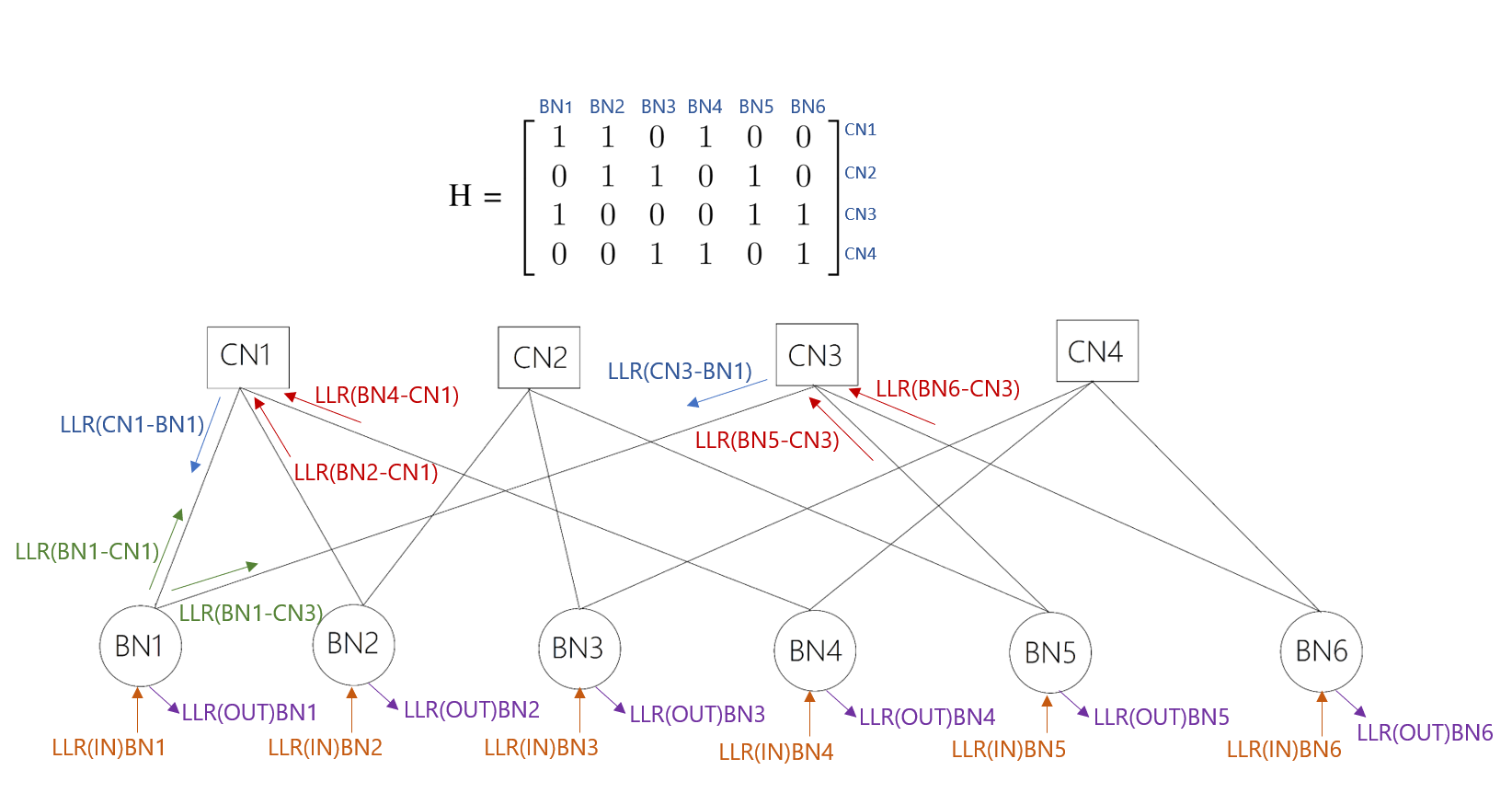}
    \caption{A parity check matrix and its Tanner Graph representation, as described in [7]}
    \label{fig:tanner}
\end{figure}
This relationship can also be visualized with a Tanner graph: where in the first row, all the bit nodes $\{b_i\}$ are placed, and in the second row all the check nodes $\{c_i\}$ are placed; every bit node has an edge to the check nodes whose check constraint it participates in, and every check node as an edge with every node that participates in its check constraint. See 
[7] and other standard references for diagrammatic view of Tanner Graph and the involved Log Likelihood Ratios (LLRs) for decoding (e.g., Fig.\eqref{fig:tanner})

Also, in the parity matrix $\mathbf{H}$, every row corresponds to a check constraint and every column indicates which bit nodes participate in the check constraint. A generator matrix $\mathbf{G}$ is obtained from the parity matrix $\mathbf{H}$. $\mathbf{G}$ is used to encode the transmitted message $u$ as,
\begin{equation}
\label{eqn:4}
    c\ =\ u\ \mathbf{G}
\end{equation}
where multiplication is mod-2. 

LDPC coded messages are decoded at a receiver using an appropriate decoding algorithm. \textit{Belief Propagation} (BP), a message-passing algorithm with numerous applications, is the most popular choice to decode LDPC coded messages.

\begin{figure*}
\begin{center}
    
    \includegraphics[width=1.5\columnwidth]{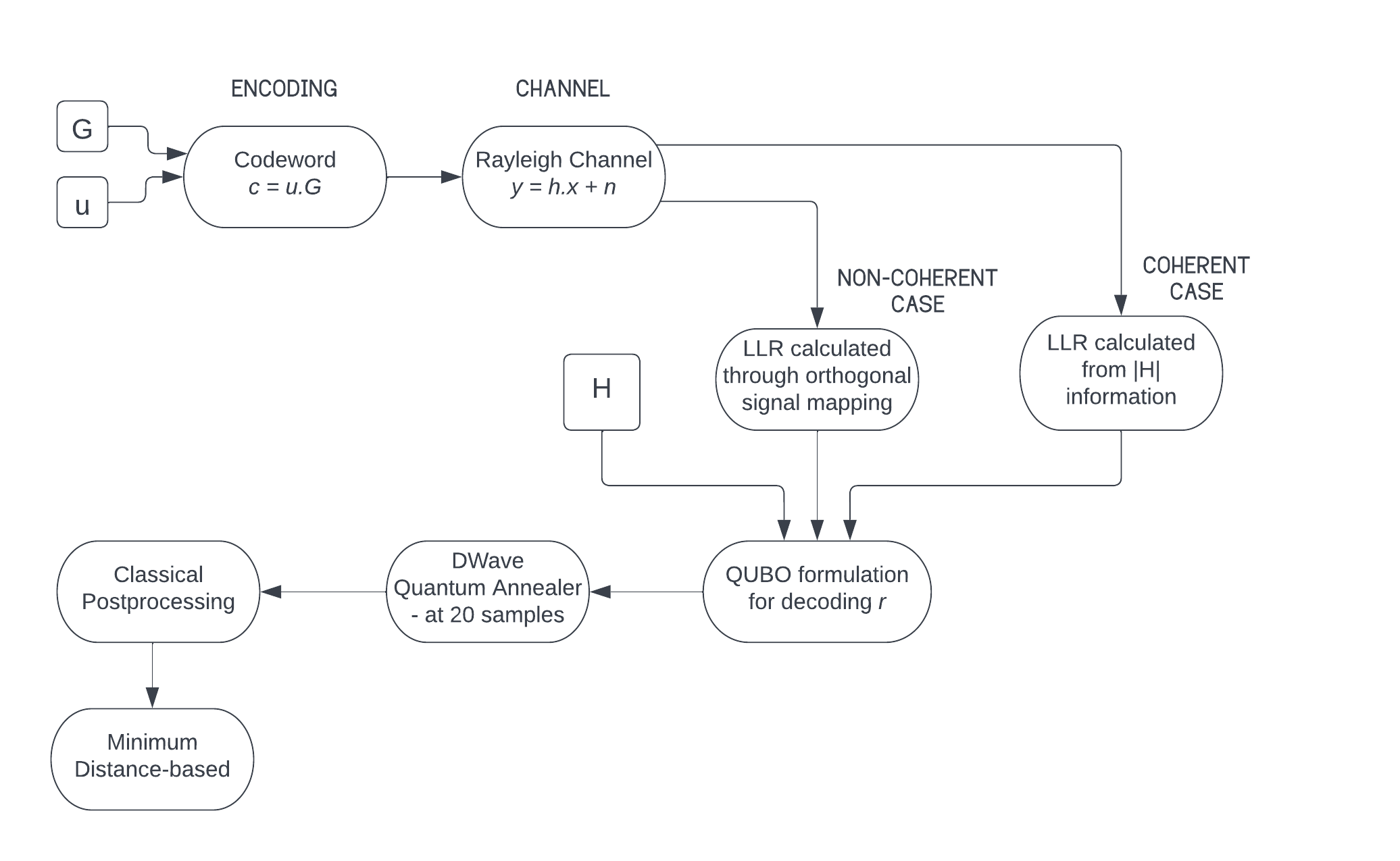}
    \caption{A schematic of the proposed approach}
    \label{fig:Flowchart}
    \end{center}
\end{figure*}

Consider the transmission of a symbol $x$. The channel model and LLR calculation are partly derived from 
[3] for the discrete-channel model. The received information $y$ can be expressed, assuming a channel gain $h$ and additive noise $n$, as,
\begin{equation}
\label{eqn:8}
    y\ =\ hx\ +\ n
\end{equation}
A Rayleigh Faded channel assumes $h \in \mathcal{C}\mathcal{N}(0, 1)$. Therefore, $h_r, h_i \sim \mathcal{N}(0, 1/2)$, where $h_r = \text{Re}\{h\}$ and $h_i = \text{Im}\{h\}$.

The following aspect is implied in the channel model: the transmitted symbol propagates through multiple paths. Each path add its own fade and phase delay. However, summing the effects of the numerous paths and obtaining the resultant at the receiver, we arrive at the the complex Gaussian model (as a result of the Central Limit Theorem).

We also assume in the Rayleigh Faded channel model, an additive white Gaussian noise (AWGN): therefore, $n \sim \mathcal{C}\mathcal{N}(0, N_0)$. $n = n_r + jn_i$ and therefore, $n_r, n_i \sim \mathcal{N}(0, N_0/2)$. The magnitude of $h$, $|h|$, is distributed according to the Rayleigh distribution (hence the name Rayleigh Faded). That is, $|h| \sim re^{-r^2/2}$, $r \geq 0$.

\subsection{Quantum Annealing}
\textit{Quantum Annealing} is a meta-heuristic optimization technique that is used to find the global minima of an objective function. A differentiating feature of Quantum Annealing over its classical counterpart,the Simulated Annealing is the presence of quantum tunnelling-type fluctuations. These fluctuations allow the optimizer to escape local minima and find global minima by "tunnelling" through barriers that separate such minima. A Quantum Annealer (QA) is an analog computer, composed of physical qubits, that implements quantum annealing. 

\par \textbf{QUBO}. Utilization of a QA requires the formulation of the target optimization problem as a Quadratic Unconstrained Binary Optimization (QUBO) problem. A QUBO problem is specified by a quadratic polynomial $f(\bar{q})$ (called the \textit{QUBO objective}), where $\bar{q} = [q_1, q_2, \dots, q_n]^{T}$, with each $q_i \in \mathbb{B} = \{0, 1\}$. Thereby, $f : \mathbb{B}^n \rightarrow \mathbb{R}$ is expressed as:
\begin{equation}
\label{eqn:1}
    f = \sum_{i} h_i q_i + \sum_{i < j} J_{i j} q_i q_j 
\end{equation}
\par The linear terms $h_i q_i$ are called the \textit{bias terms} of the $f_Q$. $h_i$ is the \textit{linear bias} of $q_i$. The quadratic terms $J_{ij} q_i q_j$ are called the \textit{coupler terms}. $J_{ij}$ is the strength of the \textit{coupler} between $q_i$ and $q_j$. An equivalent \textit{Ising form} also exists for every QUBO problem and can be obtained with the transformation for each $q_i$, $x_i:= 2q_i - 1$. The new polynomial so obtained $g(x_1, \dots, x_n)$ is equal to $f$, but its variables $x_i$ lie in $\{-1, +1\}$. The QUBO problem involves finding $\bar{q^*} \in \mathbb{B}^{n}$ such that
\begin{equation}
\label{eqn:2}
    \bar{q^*} = \underset{\bar{q} \in \mathbb{B}^n}{\text{argmin}} f(\bar{q})
\end{equation}
\par The objective is mapped to the qubits of the QA by a process known as \textit{embedding}. The qubits of the QA thereby form an \textit{Ising system}. By converging through quantum annealing, to the ground state of this Ising system, the QA is able to solve the embedded QUBO problem. The  QA is designed to "naturally" solve the optimization, by seeking the natural low-energy state of the physical system. In certain scenarios, quantum annealing can be more efficient than a classical implementation for solving QUBO problems, at least in principle. Recently, QAs have been commercialized, most notably by DWave Systems Inc. Current offerings by DWave include DWave 2000Q which houses more than 2000 qubits, and DWave Advantage, which houses more than 5000 qubits, at the time of writing.




\subsection{Quantum Belief Propagation}
In the paper by Kasi and Jamieson 
[6], they introduce a quantum version of the BP algorithm called \textit{Quantum Belief Propagation} (QBP) and designed the algorithm for implementation on quantum annealers. As discussed in the section on quantum annealing, utilizing a quantum annealer requires the formulation of the problem as a QUBO problem. Therefore, the LDPC decoding problem is formulated as a QUBO problem. The QUBO objective function comprises two terms: an \textit{LDPC satisfier} function, $\sum_{\forall c_i \in V} L_{sat}(c_i)$, to prioritize solutions that satisfy the LDPC check constraints ($L_{sat}(c_i) = 0$ if check constraint for check node $c_i$ is satisfied), and a \textit{distance} function $\sum_{i=0}^{N} \Delta_i$, to evaluate the closeness of the solutions to the received information. A low value for both these quantities are desirable. The entire QUBO function is therefore a weighted combination of these two quantities. Hence, the QUBO problem is expressed as:
\begin{equation}
\label{eqn:5}
    \underset{q_i}{\text{min}} \biggl\{ W_1 \sum_{\forall c_i \in V} L_{sat}(c_i) + W_2 \sum_{i=0}^{N} \Delta_i \biggr\}
\end{equation}

\par Let $q_1, q_2, \dots, q_n$ be the qubit variables in the QUBO objective corresponding to the sent information $x_1, x_2, \dots, x_n$. $q_i$ yields an estimate for $x_i$. This estimate is obtained by appropriately processing the received information $y_1, y_2, \dots, y_n$. 
\par \textbf{LDPC satisfier function.} The LDPC satisfier function at check node $c_i$ is $0$ if the check constraint at $c_i$ is satisfied. The only constraint for LDPC encoding is the \textit{modulo-2 sum} constraint. That is, the constraint is satisfied if the modulo-2 sum at $c_i$ is $0$. Hence, $L_{sat}$ is defined as:
\begin{equation}
\label{eqn:6}
    L_{sat}(c_i) = \Big(\big(\Sigma_{\forall j:h_{ij}=1}q_j\big) - 2L_{e}(c_i)\Big)^2 
\end{equation}
where $L_e(c_i)$ is defined in terms of some additional ancilliary qubits $\{q_{e_i}\}$. The $2L_e(c_i)$ in $L_{sat}(c_i)$ represents an even number. If there exists an even number $2L_e(c_i)$ such that the check-sum at $c_i$ equals it, then the constraint is satisfied, and expectedly $L_{sat}(c_i)$ is $0$.
\par \textbf{Distance function.} A distance $\Delta_i$ is defined that computes the proximity of the received bit information $y_i$ with the qubit $q_i$:
\begin{equation}
\label{eqn:7}
    \Delta_i = (q_i - Pr(q_i = 1|y_i))^2
\end{equation}
The probabilities $Pr(q_i = 1|y_i)$ can be obtained from the \textit{log-likelihood ratios} (LLR) for the channel noise assumed. 


\section{QUBO Formulation for Rayleigh Channels}
In our earlier work 
[10] [11], we carried out studies related LDPC decoding for AWGN channel, brought into the context the classical post processing for improved performance.
In the present work, we extend the pipeline to Rayleigh Fading channel scenario, which to the best of our knowledge is not examined within the Quantum-Enhanced Processing. Again, to reiterate, to the best of our knowledge, the extension of quantum enhanced LDPC decoding to Rayleigh fading scenario augmented with classical post processing does not exist elsewhere.

For evaluating the performance of our simulation experiment, we have used BER as the primary metric for received signal quality. Bit Error Rate (BER) is the percentage of bits that have errors relative to the total number of bits received in a transmission, usually expressed as ten to a negative power. Frame Error Rate (FER) is the ratio of data received with errors to total data received and is used to determine the quality of a signal connection. 

\par In this section, we describe our method for decoding the LDPC coded message, from the received information. We assume propagation of the transmitted message through a Rayleigh channel. We consider the two cases of channel-state knowledge: coherent and non-coherent. We compare the decoding performance in a range of noise levels. We have given a brief outline of our strategy in form of a flowchart in Fig. \eqref{fig:Flowchart}.

We assume transmission of Binary Phase-shift keying (BPSK) symbols. Let $x_1, x_2, \dots, x_n$ be the transmitted symbols. Then $x_i \in \{-1, +1\}$. Channel is Rayleigh, therefore, $h_i \sim \mathcal{C}\mathcal{N}(0, 1)$, and $n_i \sim \mathcal{N}(0, N_0)$. At the receiver, at time instant $i$, we receive $y_i = h_i x_i + n_i$. To construct the QUBO objective for the QBP, we require the parity matrix $\mathbf{H}$ that is used in the LDPC scheme, to construct the LDPC satisfier function. To construct the distance function, we require the for every $i$, $p_i = Pr(q_i = 1|y_i)$. Recall that the $q_i$ represents the sent symbol $x_i$. This can be computed from the log-likelihood ratio (LLR) 
\begin{equation}
    \label{eqn:9}
    l_i = \log\frac{Pr(x_i = +1|y_i)}{Pr(x_i = -1|y_i)}
\end{equation}
where $p_i$ is calculated as,
\begin{equation}
\label{eqn:10}
    p_i\ = \ \frac{(e^{l_i} - 1)}{(e^{l_i} + 1)}
\end{equation}

The two cases of coherent and non-coherent are put across in the following. These two cases are considered in a purely classical setting for approximating the LLRs in 
[3].
\subsection{Coherent}
In the coherent case, we assume that the channel is known to perfection at the receiver 
[12]. With $h_i$ at the receiver, a sufficient statistic 
\begin{equation}
\label{eqn:11}
r_i = \text{Re}\bigg\{\big(\frac{h_i}{|h_i|}\big)^{*}y_i\bigg\} = |h_i|x_i + z_i
\end{equation}
is computed, where $|h_i| \sim re^{r^2/2}$ and $z_i \sim \mathcal{N}(0, N_0/2)$; \textit{r} is an instance of the random variable corresponding to envelope of the channel response. We obtain $l_i$ as,
\begin{equation}
\label{eqn:11}
    l_i = 2|h_i|y_i
\end{equation}
From $l_i$, we obtain $p_i = p(q_i = 1|r_i) = \frac{(e^{l_i} - 1)}{(e^{l_i} + 1)}$. These quantities, $p_i, i = 1, \dots, n$ are used to construct the QUBO objective.

\subsection{Non-coherent}
In the non-coherent case, we do not have any information about the channel at the receiver 
[12]. To make the decoding in this case possible, we follow an \textit{orthogonal mapping scheme} for transmission. In this scheme, the message symbols $x \in \{-1, +1\}$ are mapped to $\mathbf{t} \in \{(0, a), (a, 0)\}$, $\mathbf{t} = (t_1, t_2)$. At the receiver, we obtain $y_1 = h_1 t_1 + n_1$ and $y_2 = h_2 t_2 + n_2$. Therefore, we obtain a 2-symbol vector $\mathbf{y} = (y_1, y_2)$. Given $h_1, h_2 \sim \mathcal{C}\mathcal{N}(0, 1)$ and $n_1, n_2 \sim \mathcal{C}\mathcal{N}(0, N_0)$, we can compute 
$l = \log{\frac{Pr(x=+1|\mathbf{y})}{Pr(x=-1|\mathbf{y})}}$ (mathematically similar to Equation. \eqref{eqn:9}) as:
\begin{equation}
\label{eqn:12}
    l = \frac{\big(|y_1|^2 - |y_2|^2\big)a^2}{(a^2 + N_0)N_0}
\end{equation}
Here we define the complex normal distribution given as, $\mathcal{C}\mathcal{N}$. In probability theory, the family of complex normal distributions, denoted $\mathcal{C}\mathcal{N}$ or $\mathcal{N}_{\mathcal{C}}$, characterizes complex random variables whose real and imaginary parts are jointly normal 
[2].

From this we obtain $p = Pr(x = 1|\mathbf{y}) = (e^l - 1)/(e^l + 1)$. Hence, for a sequence of message symbols $\{x_i\}$, the corresponding $\{p_i\}$ can be computed and the QUBO objective is constructed.
\begin{figure*}
\begin{center}
\begin{subfigure}{0.49\textwidth}
    \includegraphics[width=.99\columnwidth]{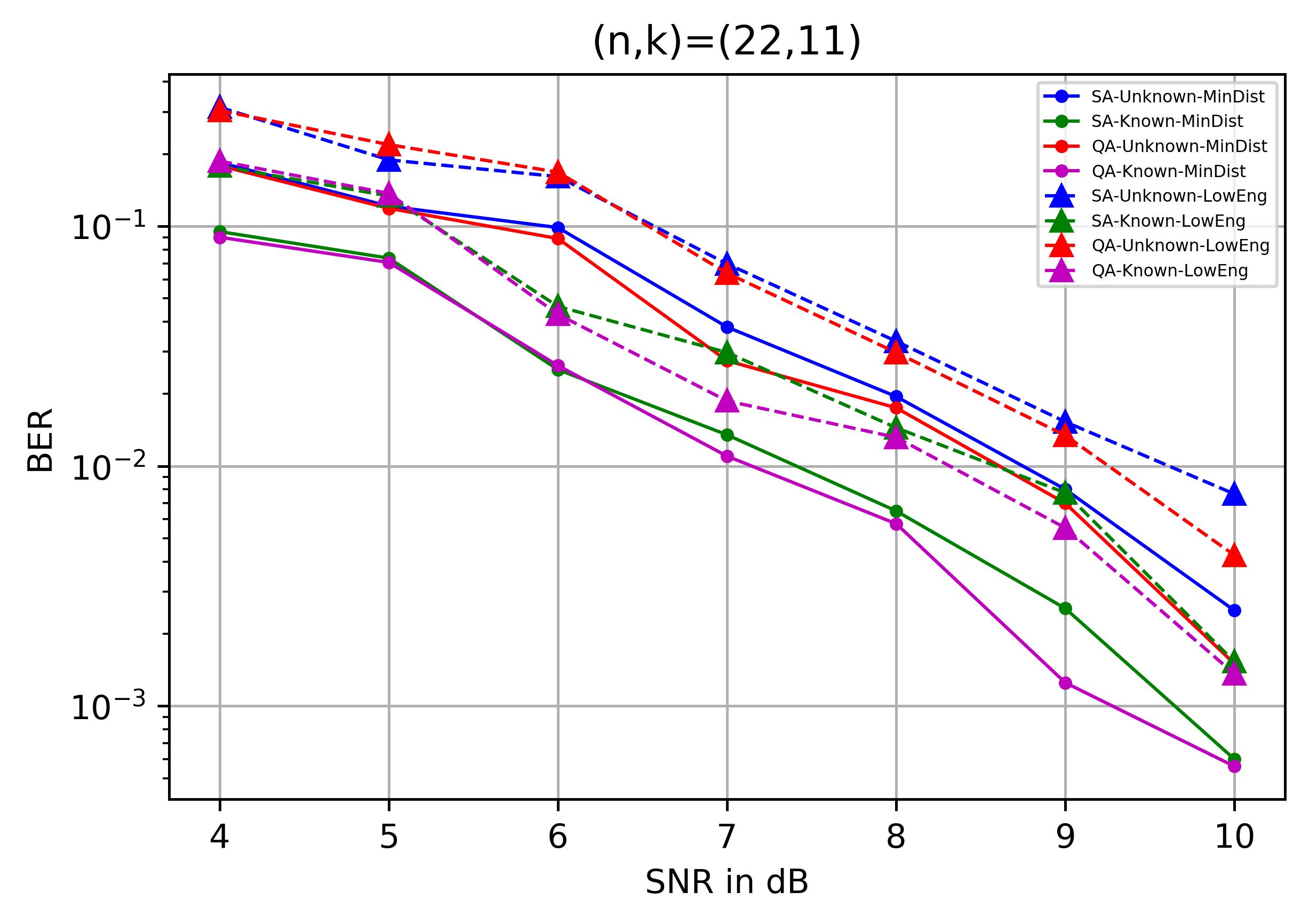}
    \caption{BER vs SNR for codeword length of (22,11) for Rayleigh case}
    \label{fig:R2211}
\end{subfigure}
\hfill
\begin{subfigure}{0.49\textwidth}
    \includegraphics[width=.99\columnwidth]{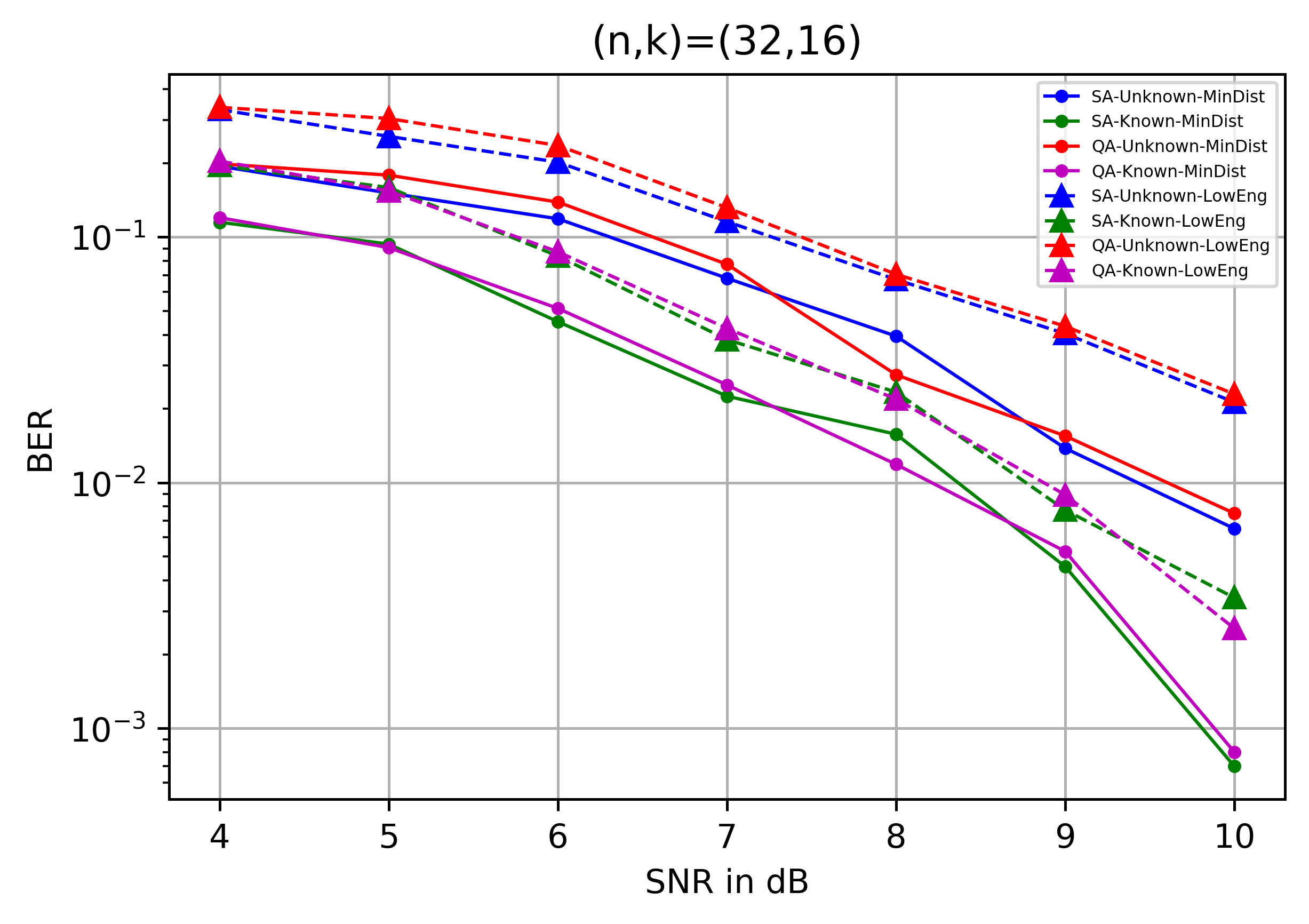}
    \caption{BER vs SNR for codeword length of (32,16) for Rayleigh case}
    \label{fig:R3216}
\end{subfigure}
\hfill
\begin{subfigure}{0.49\textwidth}
    \includegraphics[width=.99\columnwidth]{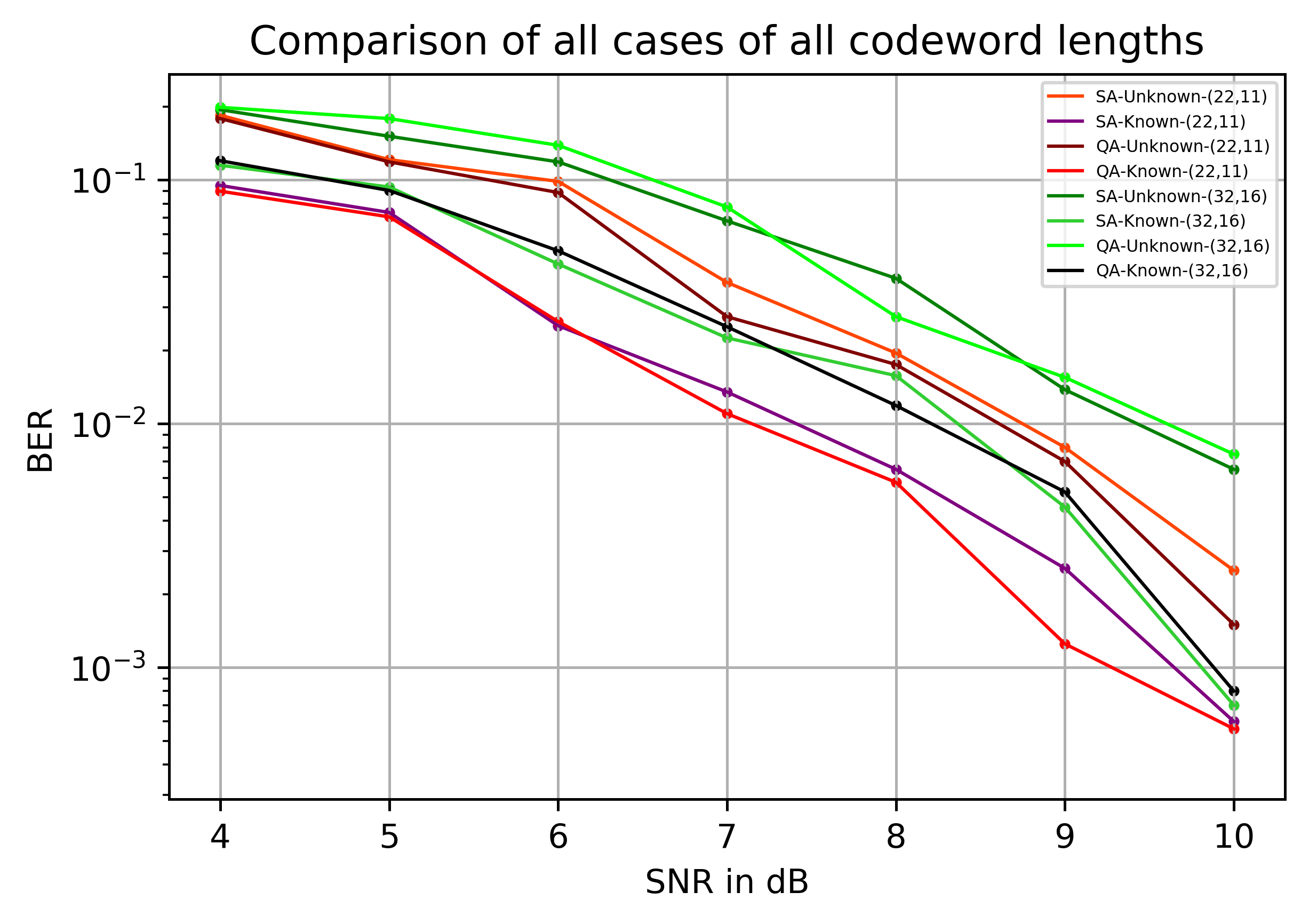}
    \caption{BER vs SNR for all codeword lengths in the Rayleigh channel case}
    \label{fig:RAll}
\end{subfigure}
\hfill
\begin{subfigure}{0.49\textwidth}
    \includegraphics[width=.99\columnwidth]{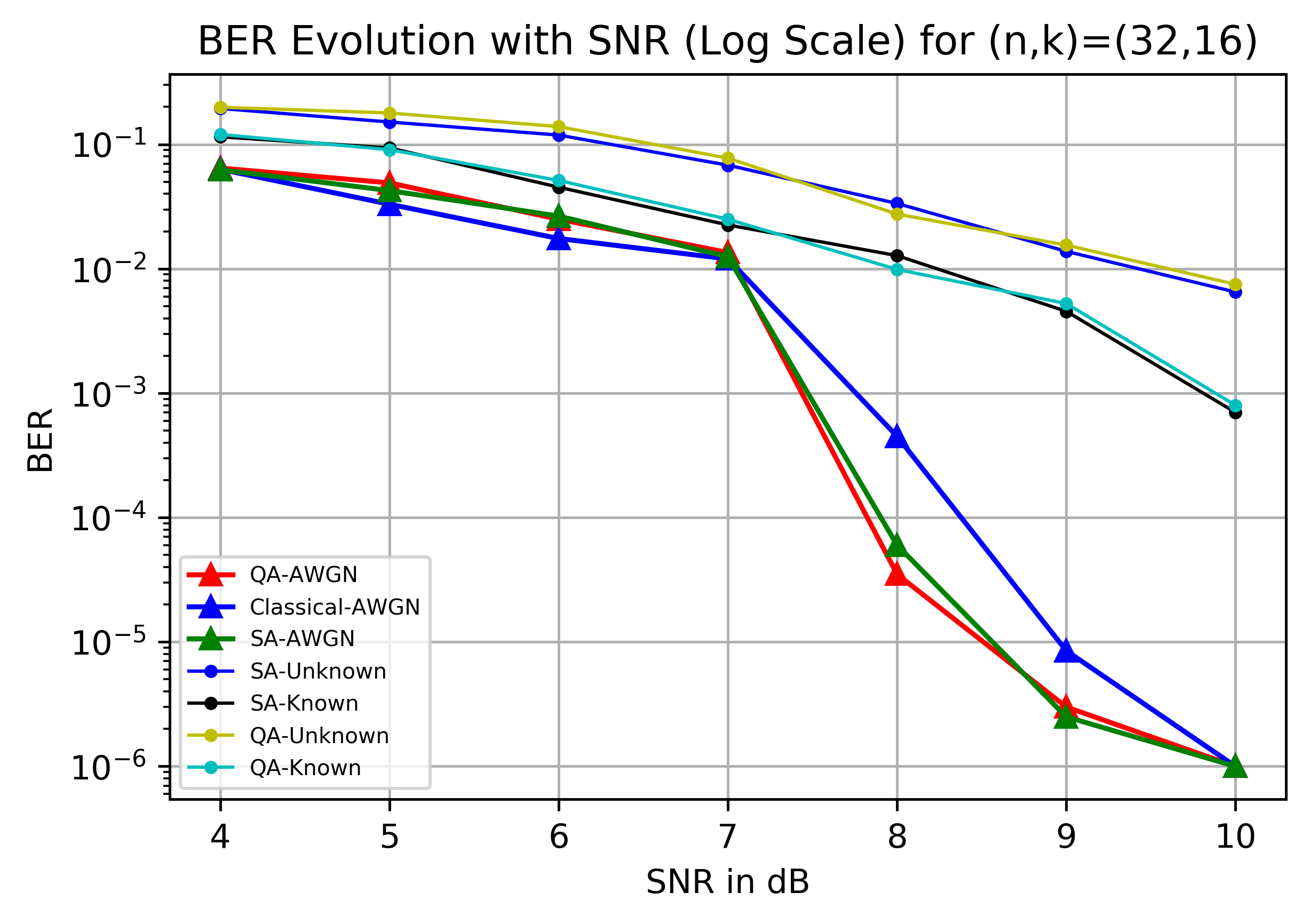}
    \caption{BER vs SNR comparison for codeword length of (32,16) between AWGN and Rayleigh Fading channels (as indicated in the legend)}
    \label{fig:RAWGN}
\end{subfigure}
\end{center}       
\caption{Results based on Bit Error Rate performance}
\label{fig:Results}
\end{figure*}

\subsection{Optimization}
The QUBO objective is input to the annealer. The annealer performs $N_a$ anneals and thereby provides $N_a$ solutions $\hat{\mathbf{x}}_{1}, \hat{\mathbf{x}}_{2}, \dots, \hat{\mathbf{x}}_{N_a}$. Due to the probabilistic nature of annealing and sampling from the low-energy state that results from the annealing, the annealer "naturally" provides a form of diversity. This notion was spelt out in our recent work 
[10] [11]; more details available there. One of the obtained solutions is highly likely to be close the transmitted message sequence $\mathbf{x} = (x_1, x_2, \dots, x_n)$. Rather than choosing the lowest energy solution from the $N_a$ solutions obtained and discarding the others, we obtain our estimate for the transmitted sequence $\hat{\mathbf{x}}$ from all the $N_a$ solutions as:
\begin{equation}
\label{eqn:14}
    \hat{\mathbf{x}} = \underset{\mathbf{x} \in \mathbb{R}^n}{\text{argmin}} \sum_{i=1}^{N_a} \lVert \hat{\mathbf{x}_i} - \mathbf{x} \rVert_{2}
\end{equation}
That is, we have carried out simple minimum distance post-processing to pick the right solution; can be linked to a kind of selection diversity reception.
We summarize again the classical processing augmented decoding: 
\subsubsection{Application on D-Wave platform}
Now, we obtain the final QUBO expression as described in \eqref{eqn:5}, which is to be optimized for the lowest energy solution first, followed by classical postprocessing.
The QUBO is thus, passed to the D-Wave quantum annealer. Several samples are collected by running the annealer multiple times. A total of 20 candidate solutions are retrieved over $10^6$ iterations, and repeated for each individual integral value of Signal-to-Noise Ratio (SNR) in the range of $[4,\ 10]$. See Fig. 1 of 
[10] [11], and the related details.

We finally check the constraints applied on the objective function, and codewords that satisfy LDPC constraints, refered to in Eq. \eqref{eqn:6}, are selected out from the samples returned from the D-Wave annealer. 

\subsubsection{Classical Postprocessing}
Finally, minimum distance decoding is performed with the received signal to obtain the final decoded codeword.

20 shots were made by the quantum annealer and the unique ground state solutions were stored. Then the solution with the minimum Euclidean distance from the received codeword was chosen as the final solution.

This classical post-processing strategy gave improved results over simply calculating the lowest energy solution returned by the annealer.

As discussed in our previous work 
[10] [11], for both simulated and quantum annealing schemes, message passing iterations are not required, as the annealer settles the solutions "naturally". The process described thus far is same for both simulated annealing and quantum annealing approaches, wherein simulated annealing runs on classical hardware with a simulated annealer sampler, and quantum annealing sampler runs on real quantum hardware. 

\section{Results}
\label{sec:Res}
As a precursor to the simulation study of Raleigh fading scenario, we produced result for a typical time-varying SNR channel in our earlier work 
[10] [11].
In this paper, we consider the actual experimentation with Rayleigh fading, where we have synthesized Rayleigh fading channels for signal transmission and performed quantum and simulated annealing-based decoding on the received signals. For our experiment we have kept the number of anneals fixed at 20, for both quantum and simulated annealing. We have conducted the experiment for $10^6$ Monte Carlo simulations and obtained the results shown in fig. \eqref{fig:Results}

For our current simulation, we have used the shape parameter of the Rayleigh distribution as 1. 

For simulated annealing, traditionally only one candidate solution is picked up.
But in order to provide a one to one correspondence with the QA based
post processing strategy, we have considered set num\_reads parameter to 20 to obtain 20 solutions of varying energies in the same annealing run and from these we eliminate the non-unique and non-constraint satisfying solutions. Then, in the remaining pool of the valid solutions, we pick that one which is the closest to the received signal vector. This strategy may be useful in fully classical processing when the Simulated Annealing meta heuristic is used. 

\subsection{Observations}
\begin{itemize}
    \item Minimum distance postprocessing is performing better than the lowest energy solutions from the solution set as can be seen from Fig. \eqref{fig:R2211} and Fig. \eqref{fig:R3216} 
     \item The BER performance curves show that the performance of Annealer-based schemes is superior to that of traditional classical belief propagation, for both Rayleigh and AWGN channels.
\end{itemize}

The result trends resemble the behaviour of received signals in a Rayleigh fading channel. The Bit Error Rates (BERs) of (32,16) codeword length signal transmitted in the coherent case, as can be seen in Fig. \eqref{fig:R3216} remain between $10^{-1}$ and $10^{-3}$ for an SNR range of 4-10. Both quantum annealing and simulated annealing have given comparable results throughout the SNR range, with their results significantly improving regardless of codeword length, in case of the minimum distance postprocessing, as compared to the lowest energy solutions from the annealer, most visibly illustrated in Fig. \eqref{fig:R2211}\eqref{fig:R3216}\eqref{fig:RAll}. Our strategy is quite similar to our earlier work in 
[10] [11], with AWGN-infested channel resulting in more accurately decoded signals than the Rayleigh fading channel, as is expected.

As pointed in our earlier work 
[10] [11], the performance of the SA and the QA are quite comparable, but, it is important to note that the QA-based results are run on the actual (noisy) hardware of the present NISQ era, and the devices are continuously evolving. With improved devices in terms of reducing different sources of noise, in future, apart from able to run larger problems (in our context, decoding of longer code lengths), we may get better performance from QA. 

\subsection{Additional Remarks on Diversity}
The notion of diversity due to operational nature of QA and exploiting it through classical post-processing is tested for couple of small code lengths, as elaborated in the paper. In order to provide a level-playing field even for fully classical SA, we considered different outputs of SA and post-processed them in a similar way, as mentioned earlier. Of course, results captured in Figure \eqref{fig:Results}, did not show up in useful improvement of QA-based scheme compared to fully classical SA-based scheme. Of course, there is an improvement over traditional work horse of classical Sum-Product or BP decoding. As pointed out in 
[10] [11], the usefulness of diversity because of Quantum Computing is suggested  based on the fact in  Quantum Computing,  the  probability  amplitudes  can  interfere  unlike  in  classical  probabilistic computing. Together with other quantum effects, in principle, one can see a different potential of Quantum compared to classical processing. The preliminary results captured here suggest that further research is needed in this direction, both experimental and theoretical. One can consider experimenting with different annealers available, including DWave Advantage. The quantum effects may be more pronounced in Gate-Model Universal Quantum Computers, and once the devices with enough number of qubits are available, experimentation can be run on them. Of course, with the improved noise and error characteristics of the Quantum Computers in the coming years, studies through them may provide better conclusions.The theoretical study of possible diversity gain, in the context we have considered, to the best of our knowledge is wide open. The whole idea is, even if some gain due to diversity is available, why not use it? It is worth noting at this juncture again that the quantum and classical computers work together in solving difficult real-life problems, not only in the NISQ era, but, even in the futuristic fault-tolerant years. Quantum Computers may solve large-sized optimization problems more comfortably and naturally in future, and utilizing the diversity, even if the gain is small in the quantum-classical pipeline, can be beneficial.
\section{Conclusion}
The paper brought out results of small-sized examples of LDPC decoding, using Quantum Annealer and Classical Computing pipeline, for Rayeligh fading channels. The post processing was limited to simple minimum-distance based selection of the candidate outputs from different runs. The performance was superior to the widely adopted classical Belief Propagation. 

The paper addressed two popular cases of fully known and unknown channel scenarios, and the work is underway to formulate the algorithm for partially known channel case and its performance. This may open up further avenue for Classical Post-Processing in the whole decoding/inference process. Apart from extending beyond simple minimum-distance decoding, the idea can be taken further to other complex baseband signal processing tasks, putting into play the tandem working of classical and quantum computing. As discussed in the last paragraph of the previous section, further study is warranted for assessing the possible diversity benefit from Quantum Processing. Studies based on simulation and hardware execution on gate based universal quantum computing also of importance as a future research thread. 

\end{document}